\documentstyle[prb,aps,multicol,epsf]{revtex}

\begin{document}
\draft

\title{Domain wall superconductivity in hybrid superconductor --
 ferromagnetic structures}
\author{A.~Yu.~Aladyshkin, $^{(1)}$ A.~I.~Buzdin, $^{(2)}$
 A.~A.~Fraerman, $^{(1)}$ A.~S.~Mel'nikov, $^{(1)}$
 D.~A.~Ryzhov, $^{(1)}$\cite{e-mail} A.~V.~Sokolov $^{(1)}$}
\address{$^{(1)}$ Institute for Physics of Microstructures,
 Russian Academy of Sciences, 603950, Nizhny Novgorod, GSP-105, Russia,\\
 $^{(2)}$ Centre de Physique Moleculaire Optique et Hertzienne,
 Universite Bordeaux 1-UMR 5798, CNRS, F-33405 Talence Cedex, France}
\date{\today}
\maketitle

\begin{abstract}
On the basis of phenomenological Ginzburg-Landau approach we investigate
the problem of order parameter nucleation in hybrid
superconductor/ferromagnetic (S/F) systems with a domain structure in
applied external magnetic field. Both the isolated domain boundaries and
periodic domain structures in ferromagnetic layers are considered. We
study the interplay between the superconductivity localized at the domain
walls and far from the walls and show that such interplay determines a
peculiar field dependence of the critical temperature $T_c$. For a
periodic domain structure the behavior of the upper critical field of
superconductivity nucleation near $T_c$ is strongly influenced by the
overlapping of the superconducting nuclei localized over different
domains.
\end{abstract}


\section{Introduction}

The problem of coexistence of superconducting and magnetic orderings has
been studied for several decades (see, e.g., Refs. \cite{general,phys} for
review). One can separate two basic mechanisms responsible for interaction
of superconducting order parameter with magnetic moments in the
ferromagnetic state: (i) the electromagnetic mechanism (interaction of
Cooper pairs with magnetic field induced by magnetic moments) which was
first discussed by V.~L. Ginzburg \cite{ginz} in 1956; (ii) the exchange
interaction of magnetic moments with electrons in Cooper pairs. The
revival of interest to the fundamental questions of magnetism and
superconductivity coexistence has been stimulated, in particular, by the
recent investigations of the hybrid superconductor/ferromagnetic (S/F)
systems. Such thin film structures consist of a ferromagnetic insulator
film and superconducting film deposited on it. Similar situation can be
obtained with a metallic ferromagnet when a superconducting film is
evaporated on the buffer oxide layer in order to avoid proximity effect.
The superconducting properties of such structures attract a growing
interest due to a large potential for applications. In particular, such
hybrid S/F systems are intensively investigated in connection with the
problem of controlled flux pinning. The enhancement of the depinning
critical current density $j_c$ has been observed experimentally for
superconducting films with arrays of submicron magnetic dots,
\cite{Martin,Morgan,Bael-2} antidots, \cite{Bael-1} and for S/F bilayers
with domain structure in ferromagnetic films. \cite{Santiago} Theory of
vortex structures and pinning in the S/F systems at rather low magnetic
fields (in the London approximation) has been developed in Refs.
\cite{sonin,iosif,Lyuks,Sasik,Bul,Besp,Erdin,Helseth,Milosevic-2,Sonin-2}.

A nonhomogeneous magnetic field distribution induced by the domain
structure in a ferromagnetic layer influences strongly the conditions of
the superconducting order parameter nucleation, and, as a consequence, the
hybrid S/F systems reveal a nontrivial phase diagram in an external
applied magnetic field ${\bf H}$ (see, e.g., Refs.
\cite{Otani,Lange-2002,lange2}). In this paper we focus on the theoretical
study of this phase diagram on the basis of phenomenological
Ginzburg-Landau (GL) model. We assume that the electromagnetic mechanism
mentioned above plays a dominant role and neglect the exchange interaction
which is obviously suppressed provided superconducting and ferromagnetic
layers are well separated by an insulating barrier. We also assume that
the domain walls are well pinned and do not take account of changes in the
domain structure with an increase in $H$.

The distribution of the magnetic field induced by the domain structure is
determined by the ratio of two length scales: thickness of ferromagnetic
film $D$ and distance between the domain walls $w$ (hereafter we neglect a
finite width of the domain wall, i.e. consider this width to be much less
than the superconducting coherence length). Provided the ferromagnetic
film is rather thick ($D\gg w$), the magnetic field in a thin
superconducting film is almost homogeneous over the domain and suppresses
the critical temperature of superconductivity nucleation. In this case
with the decrease in the temperature the superconductivity must firstly
appear just above the domain wall (see Refs. \cite{buzdin,buzdin2}) due to
the mechanism analogous to the one responsible for the surface
superconductivity below $H_{c3}$ (see Ref. \cite{Saint-James}). Thus, in
this limit the domain walls stimulate the nucleation of the
superconducting order parameter. Note, that the same effect should reveal
for two-dimensional magnetic field distributions induced, e.g., by
magnetic dots, and results in the dependence of the upper critical field
on the angular momentum of the superconducting nucleus wave function (see
Refs. \cite{2d1,2d2,2d3}).

For a thin ferromagnetic film ($D\ll w$) the magnetic field decays with
the increase in the distance from the domain wall and almost vanishes
inside the domain. In the absence of the external field such domain wall
should locally weaken superconductivity as it was discussed in Ref.
\cite{sonin}. The superconducting nucleus in this case should appear far
from the domain wall. As we switch on an external magnetic field, we can
control the position of the superconducting nucleus suppressing the order
parameter inside the domains. Thus, the phase diagram of the S/F bilayer
is generally determined by the interplay between the superconductivity
nucleated at the domain walls and in between these walls. For small period
domain structures (when $w$ is comparable with the nucleus size) this
simple physical picture based on consideration of isolated superconducting
nuclei should be modified taking account of the interaction between the
superconducting nuclei localized above different domain walls.

\begin{figure}[htb]
 \begin{center}
 \epsfxsize=85mm
 \epsfbox{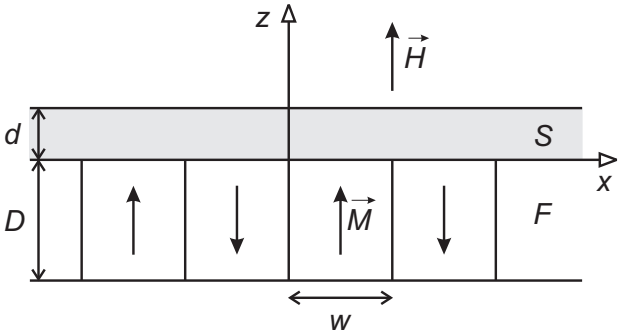}
 \end{center}
 \caption{
 \label{fig1}
Superconductor/ferromagnetic (S/F) bilayer.}
\end{figure}

Our further consideration is based on the linearized GL equation for the
order parameter $\Psi$:
\begin{equation}
 \label{GL-vec}
 - \left(\nabla+\frac{2\pi i}{\Phi_0}{\bf A}\right)^2\Psi
 = \frac{1}{\xi^2(T)}\Psi.
\end{equation}
Here ${\bf A}({\bf r})$ is the vector potential, ${\bf B}({\bf
r})=\nabla\times{\bf A}({\bf r})$, $\Phi_0$ is the flux quantum,
$\xi(T)=\xi_0/\sqrt{1-T/T_{c0}}$ is the coherence length, and $T_{c0}$ is
the critical temperature of the bulk superconductor at $B=0$. For
superconducting films with thickness $d$ much smaller than coherence
length the role of the parallel component of the magnetic field is
negligibly small. Thus, we can take account only of the magnetic field
component $B_z$ perpendicular to the film surface and also neglect the
dependence of the order parameter on $z$. For the sake of simplicity we
restrict ourselves to the consideration of the one-dimensional case:
$B_z(x)=H+b(x)$, where $H$ is a uniform external magnetic field and $b(x)$
is the $z$--component of the field induced by the magnetization ${\bf
M}=M(x){\bf z}_0$ (see Fig. \ref{fig1}).

Choosing the gauge ${\bf A}=A(x){\bf y}_0$, one can easily see that the
momentum along the $y$ axis is conserved, hence we can find the solution
of the Schr\"odinger-like equation (\ref{GL-vec}) in the form $\Psi({\bf
r})=f_k(x)\exp(-ik y)$, where function $f_k(x)$ should be determined from
a solution of the one-dimensional problem:
\begin{equation}
 \label{GL-x}
 - \frac{d^2 f_{k}}{d x^2}+\left(\frac{2\pi}{\Phi_0}A(x)-k\right)^2 f_{k}
 = \frac{1}{\xi^2(T)}f_{k} \ .
\end{equation}

Nontrivial solutions of Eq.(\ref{GL-x}) exist only for a discrete set of
temperatures $T_n(k)$. The superconducting critical temperature $T_c$
should be define as the highest value $\max\{T_n(k)\}$, corresponding to
the lowest "energy level" $1/\xi^2(T)$ of the Schr\"odinger-like equation
(\ref{GL-x}).

\section{Superconductivity nucleation at a domain wall: an isolated
order parameter nucleus}

Let us start from consideration of a superconducting nucleus at a single
domain wall taking the magnetization ${\bf M}$ near the wall in the form:
${\bf M}=M{\rm sign}(x){\bf z}_0$ (we assume that the domain wall width is
much less than the superconducting coherence length).

\subsection{Domain wall in a thick ferromagnetic film:
 step-like magnetic field profile}
 \label{2a}

As it was mentioned above, for a rather thick ferromagnetic film ($D\gg
w$) the expression for the distribution of magnetic field near the surface
reads: $B_z= 4\pi M{\rm sign}(x)+H$, where $H$ is an external applied
magnetic field. We choose the gauge in the form: ${\bf A}=(4\pi
M|x|+Hx){\bf y}_0$. At high temperatures the superconductivity far from
the domain wall can be completely suppressed due to the orbital effect. On
the contrary, near the boundary the superconducting nucleus can be still
energetically favorable due to the mechanism analogous to the one
responsible for the existence of $H_{c3}$ critical field for
superconducting nucleus near the superconductor-insulator interface (see,
e.g., Ref. \cite{Saint-James}). Thus, a change of the magnetization
direction which occurs at a domain boundary is responsible for a partial
decrease of the orbital effect which provides conditions for the formation
of localized superconducting nuclei at the domain walls at high
temperatures (above the critical temperature far from the walls). Such a
localized nucleus can appear only if we take account of proximity effect,
i.e. consider the Cooper pairs to exist on both sides of the domain
boundary. Such systems can reveal an interesting behavior in an external
magnetic field. An external magnetic field applied to the sample results
in a partial compensation of the field above one of the domains. As a
result, the critical temperature of superconductor can depend
nonmonotoneously on the applied magnetic field. Both the critical
temperature of superconductivity nucleation far from the domain wall and
critical temperature of formation of localized superconductivity at the
wall should increase up to the external field value equal to the magnetic
induction induced by the ferromagnetic moment.

It is convenient to rewrite equation (\ref{GL-x}) in the following
dimensionless form:

\begin{equation}
\label{gl3}
 -\frac{\partial^2 f_k}{\partial t^2} + (|t|+ht-t_0)^2f_k
 = Ef_k \ ,
\end{equation}
where $t=x/L$, $t_{0}=k L$, $L^{2}=\Phi_{0}/(2\pi B_0)$, $h=H/B_0$,
$E=(T_{c0}-T)/\Delta T_{c}^{orb}$, the value $\Delta
T_{c}^{orb}=T_{c0}\xi_0^2/L^2$ characterizes the shift of critical
temperature due to the orbital mechanism, and $B_0$ is the maximum
absolute value of the field $b$ (in this subsection $B_0=4\pi M$).

For the case $|t_{0}|\rightarrow\infty$ a superconducting nucleus will
appear far from the domain boundary at a certain $T_c^\infty$. In this
limit the lowest eigenvalue $E=|1-|h||$ of equation (\ref{gl3}) and,
hence, the critical temperature is not disturbed by the presence of the
domain boundary. On the contrary, for finite $t_{0}$ values the
superconducting nuclei to the left and to the right from the domain wall
can not be considered separately due to the proximity effect. Provided the
lowest energy level in the resulting potential well in equation
(\ref{gl3}) is minimal for a certain finite $t_{0} $ coordinate, we get a
superconducting nucleus localized at the domain boundary for temperatures
above $T_c^\infty$. The mechanism resulting in the appearance of such
localized nucleus is analogous to the one responsible for existence of the
surface superconductivity at the superconductor/insulator boundary for
magnetic fields $H_{c2}<H<H_{c3}$. Indeed, for $h=0$ the potential well
$V(t)$ in Schr\"odinger equation (\ref{gl3}) is symmetric ($V(t)=V(-t)$)
and the eigenvalue problem (\ref{gl3}) can be considered only for $t>0$
with the boundary condition $f_k'(t=0)=0$. For this particular case the
energy minimum corresponds to $t_0^{2}=E_{min}=0.59010$ (Ref.
\cite{Saint-James}). An increase in the $h$ value will obviously result in
an increasing asymmetry of the well $V(t)$, and, thus, in the suppression
of superconductivity localized at the domain wall. The equation
(\ref{gl3}) can be solved exactly in terms of Weber functions (see Ref.
\cite{Saint-James,math}):
\begin{equation}
 f_k=C_1 W\left(\sqrt{1+h}t-\frac{t_0}{\sqrt{1+h}}, \frac{E}{1+h}
 \right) \ , \qquad t>0 \ ,
\end{equation}
\begin{equation}
 f_k=C_2 W\left(-\sqrt{1-h}t-\frac{t_0}{\sqrt{1-h}}, \frac{E}{1-h}
 \right) \ , \qquad t<0 \ .
\end{equation}
Here $C_1$ and $C_2$ are constants, and the Weber function
$W(s,\varepsilon)$ is the solution of the following equation:
\begin{equation}
 \label{gl4}
 -\frac{\partial ^{2}W}{\partial s^{2}}+s^{2}W=\varepsilon W
\end{equation}
with the boundary condition $W(s\rightarrow
+\infty,\varepsilon)\rightarrow 0$. Matching these solutions at $t=0$ we
obtain:

\begin{equation}
 \frac{\sqrt{1+h}W_{s}^{^{\prime}} \left(-\frac{t_0}{\sqrt{1+h}},
 \frac{E}{1+h}\right)}
 {W\left(-\frac{t_0}{\sqrt{1+h}},\frac{E}{1+h}\right)}
 = -\frac{\sqrt{1-h}W_{s}^{^{\prime}} \left(-\frac{t_0}{\sqrt{1-h}},
 \frac{E}{1-h}\right)}
 {W\left(-\frac{t_0}{\sqrt{1-h}},\frac{E}{1-h}\right)}.
\end{equation}

\begin{figure}[htb]
 \begin{center}
 \epsfxsize=85mm
 \epsfbox{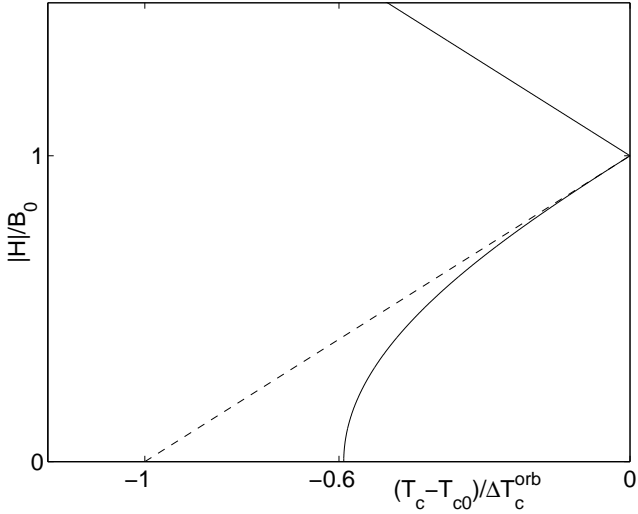}
 \end{center}
 \caption{\label{fig2}
The temperature dependence of the critical magnetic field for a S/F system
with a thick ferromagnetic layer. The solid (dashed) line corresponds to
the superconductivity nucleation at the domain boundary (far from the
domain boundary).}
\end{figure}

This equation can be solved numerically which allows us to obtain the
function $E(t_0,h)$. The resulting dependence of the critical temperature
of superconductivity nucleation on parameter $h$ is shown in
Fig.\ref{fig2}.

One can see that the external field suppresses the localized
superconducting nuclei and the superconductivity localized at the domain
wall exists only at a relatively weak applied field. As we increase an
external field the superconducting nucleus shifts away from the domain
wall towards the region where the absolute value of the total magnetic
field is minimal. For $0\leq |h|\leq 1$ the curve $E(h)$ calculated
numerically can be fitted by the following simple expression:
\begin{equation}
 \label{Appr}
 E(h) \simeq \left(E_{min}-\frac{1}{2}\right)h^4
 + \left(\frac{1}{2}-2E_{min}\right)h^2 + E_{min}.
\end{equation}

\subsection{Domain wall in a thin ferromagnetic film}
\label{model}

In this subsection we proceed with consideration of another limiting case
$D\ll w$ and consider the problem of superconductivity nucleation in the
field of an isolated domain wall in a thin ferromagnetic film:
$B_z(x,z=0)=4M\tan^{-1}(D/x)+H$. Obviously, for rather weak external
magnetic fields $H<B_0$ (in this subsection the maximum value of the
domain wall field at $z=0$ is given by the expression $B_0=2\pi M$) the
superconducting order parameter nucleates in the region near the point
$x_0$ where $B_z(x_0)=0$. Provided the localization length $\ell$ of the
superconducting nucleus is much smaller than the characteristic length
scale of magnetic field distribution, we can expand vector potential as
 $$
 A(x)\simeq A(x_0)+\frac{1}{2}B_z'(x_0)(x-x_0)^2.
 $$
Such a local approximation is valid if the following conditions are
fulfilled:
\begin{equation}
 \left|\frac{B_z''(x_0)}{B_z'(x_0)}\ell\right| \ll 1
 \qquad {\rm and} \qquad
 \ell\ll x_0.
 \label{condition}
\end{equation}
Introducing a new coordinate $t=(x-x_0)/\ell$ we obtain the dimensionless
equation
\begin{equation}
 \label{Dim-less}
 -\frac{d^2f}{dt^2} + (t^2-Q)^2f = \epsilon f \ ,
\end{equation}
\begin{equation}
 \ell = \sqrt[3]{\frac{\Phi_0}{\pi|B_z'(x_0)|}}
 = D\sqrt[3]{\frac{\Phi_0}{4\pi MD^2\sin^2(H/4M)}} \ ,
\end{equation}
\begin{equation}
 \label{Energy}
 \epsilon = \frac{\ell^2}{\xi_0^2}\left(1-\frac{T}{T_{c0}}\right) \ ,
\end{equation}
\begin{equation}
 Q = \sqrt[3]
 {\frac{\Phi_0}{\pi B_z'(x_0)}}\left(k-\frac{2\pi}{\Phi_0}A(x_0)\right).
\end{equation}

The lowest eigenvalue of Eq.(\ref{Dim-less}) $\epsilon_0\simeq 0.904$ is
achieved at $Q\simeq0.437$. For the critical temperature $T_c$ of
superconductivity nucleation we obtain:

\begin{equation}
 \label{analyt}
 \frac{T_{c0}-T_c}{\Delta T_c^{orb}}=\frac{\epsilon_0}{\pi}
 \left(\frac{\Phi_0}{2B_0 D^2}\right)^{1/3}
 \sin^{4/3}\left(\frac{\pi|H|}{2B_0}\right) \ .
\end{equation}

This expression is valid when
\begin{equation}
 \left| \frac{\sin^{1/3}(H/4M)}{\cos(H/4M)} \right|
 \ll \frac{4\pi MD^2}{\Phi_0} \ .
\end{equation}

Note that close to $T_{c0}$ the upper critical field has an unusual
temperature dependence: $\propto (T_{c0}-T)^{3/4}$.

\begin{figure}[htb]
 \begin{center}
 \epsfxsize=85mm
 \epsfbox{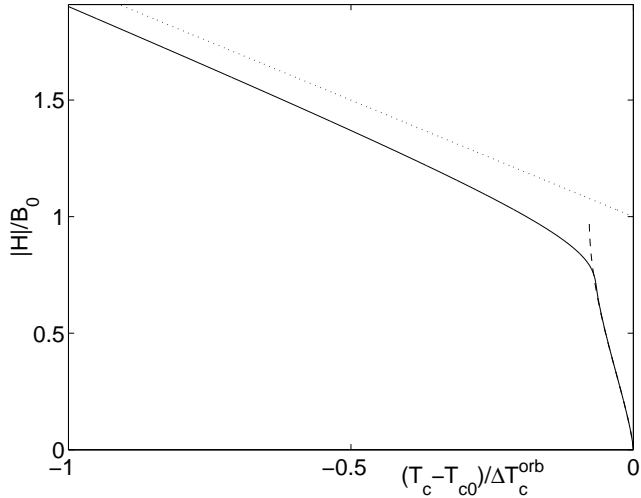}
 \end{center}
 \caption{\label{fig3}
The temperature dependence of the upper critical field for a domain wall
in a S/F system with a thin ferromagnetic layer for $B_0D^2/\Phi_0=25$
(solid line). The dash line corresponds to the analytical expression
(\ref{analyt}) at low fields, the dotted line corresponds to the high
field asymptotics (\ref{high}).}
\end{figure}

As we increase an external magnetic field $H$ the position of
superconducting nucleus shifts from infinity to the domain wall at $x=0$.
For rather large fields $H$ the nucleus appears to be localized at the
domain wall. Thus, the behavior of the nucleus coordinate in an external
field is an opposite to the one considered in the subsection \ref{2a}. The
critical temperature for high field $H$ limit is given by the expression
\begin{equation}
 \label{high}
 \frac{T_{c}-T_{c0}}{\Delta T^{orb}_c}= 1-|H|/B_0 .
\end{equation}
The simple asymptotical formulas given above are in a good agreement with
our numerical simulations of Eq.(\ref{GL-x}) (see Fig.\ref{fig3}).

\begin{figure}[htb]
 \begin{center}
 \epsfxsize=85mm
 \epsfbox{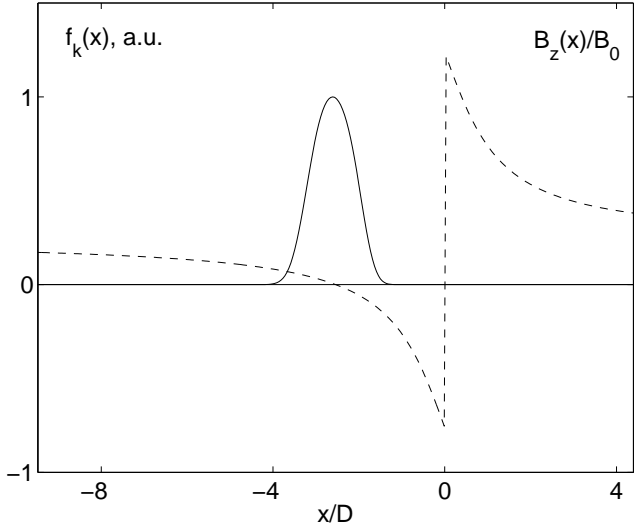}
 \end{center}
 \caption{\label{fig4}
The typical behavior of the ground state wavefunction for a domain wall in
a S/F system with a thin ferromagnetic layer (solid line). The magnetic
field profile is shown by the dash line. The parameters are
$B_0D^2/\Phi_0=25$ and $H/B_0=0.24$.}
\end{figure}

For numerical analysis of the localized states of Schr\"odinger-like
equation (\ref{GL-x}) with an external magnetic field we approximated it
on a equidistant grid and obtained the eigenfunctions $f_k(x)$ and
eigenvalues $1/\xi^2(T)$ by the diagonalization method of tridiagonal
difference scheme. The typical behavior of the ground state wavefunction
is shown in Fig.\ref{fig4}.

\section{Nucleation of superconductivity for a periodic domain structure}

In this section we consider the effect of interaction of Cooper pair
wavefunctions nucleated at different domain walls. Surely such interaction
is important only for temperatures close to $T_{c0}$ ($\xi(T)>w$),
otherwize for a rather large domain size $w\gg\xi(T)$ the overlapping of
superconducting nuclei above different domain walls is exponentially
small. For the sake of simplicity we consider here the case $w\ll D$ and
take the step-like distribution of magnetic field induced by the domain
structure with the period $a=2w$: $b(x)=B_0{\rm sign}(x)$, for $|x|<w$ and
$b(x+na)=b(x)$, where $n$ is an integer. The corresponding vector
potential can be chosen in the form: $A(x)=B_0 |x|$ for $|x|<w$ and
$A(x+na)=A(x)$.

In the absence of external field the general solution of Eq.(\ref{GL-x})
meets the Bloch theorem:
\begin{equation}
 \label{bloch}
 f_{kq}(x+a)=f_{kq}(x) e^{iq a} \ ,
\end{equation}
where $q$ is a quasimomentum. The nodeless wavefunction of the ground
state corresponds to the value $q=0$ and is an even function of $x$, and,
thus, we obtain $f_k^\prime (0)=f_k^\prime (w)=0$. So we conclude that the
solution at zero external field is identical to the one describing the
superconductivity nucleation in a superconducting film of the thickness
$w$ in the uniform magnetic field $B_0$. Following Ref. \cite{Saint-James}
we can obtain the ground state wavefunctions and energy
$E=(T_{c0}-T_c)/\Delta T_{c}^{orb}$ as a function of the momentum $k$. The
behavior of the resulting dependence of $E(k)$ strongly depends on the
parameter $w/L$. Two different regimes could be realized: for small values
of $w/L<2.5$ there is only one minimum of $E(k)$ at $k=w/(2L^2)$, which
corresponds to the superconductivity nucleation above the domain center.
For larger values of $w/L$ one obtains two minima with equal energies $E$
at $k^{min}_{1}$ and $k^{min}_{2}$ ($k^{min}_{1}+k^{min}_{2}=w/L^2$). For
$w/L \gg 1$ the coordinates of these minima $k^{min}_{1}$ and
$k^{min}_{2}$ and minimum energy $E$ approach the values corresponding to
the ones for isolated domain walls (see subsection \ref{2a}). Depending on
the $k$-momentum value the superconducting nuclei appear either above the
walls at $x=na$ (for $k^{min}_{1}$) or at $x=w+na$ (for $k^{min}_{2}$).
Thus, the nuclei at neighboring domain walls do not interact within the
linearized GL theory. The dependence of the critical temperature on the
field $B_0$ in a periodic domain structure is described by the formula:
 $$
 1-T_c/T_{c0}=\frac{4\xi^2_0}{w^2}
 F\left(\frac{\pi B_0w^2}{2\Phi_0}\right) \ ,
 $$
where the function $F(z)$ coincides with that for a superconducting film
in the uniform magnetic field $B_0$ which is plotted, e.g., in Ref.
\cite{Saint-James}. For a finite domain thickness $w$ the critical
temperature $T_c(B_0)$ appears to be larger than the one for a single
domain wall. This difference in $T_c$ becomes rather large for small
values $z=\frac{\pi B_0w^2}{2\Phi_0}$ when the nucleus is not localized
near the domain boundary. For large $z$ values one can obtain:
$F(z)\rightarrow z/1.69$, which corresponds to the dependence $T_c(B_0)$
for a nucleus at a single domain wall.

\begin{figure}[htb]
 \begin{center}
 \epsfxsize=85mm
 \epsfbox{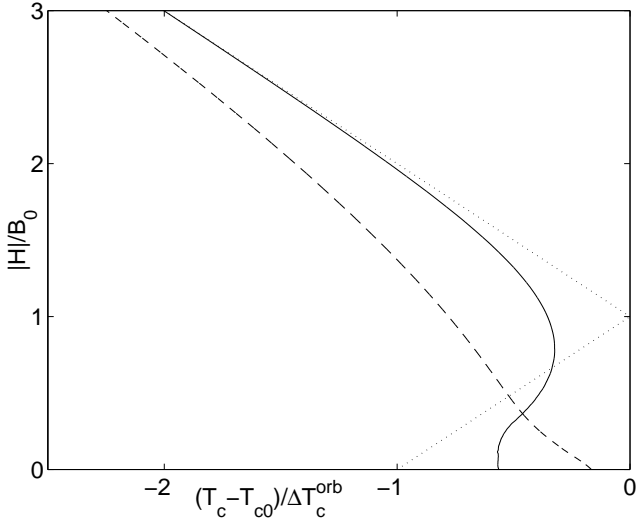}
 \end{center}
 \caption{\label{fig5}
The temperature dependence of the upper critical field for a periodic
domain structure in a S/F system with a thick ferromagnetic layer for $\pi
B_0w^2/\Phi_0=5$ (solid line) and $\pi B_0w^2/\Phi_0=1$ (dash line).}
\end{figure}

If we apply an external magnetic field $H$ the Bloch theorem is no more
valid and the solution $f_k(x)$ appears to be localized. The energy level
$E(k)$ becomes a periodic function of the momentum $k$: $E(k+4\pi H
w/\Phi_0)=E(k)$. The behavior of the upper critical field and structure of
superconducting nuclei are controlled by the parameter $w/L$. The results
of our calculations carried out using the same numerical scheme as in
subsection \ref{model} are shown in Fig.\ref{fig5}.

\begin{figure}[htb]
 \begin{center}
 \epsfxsize=85mm
 \epsfbox{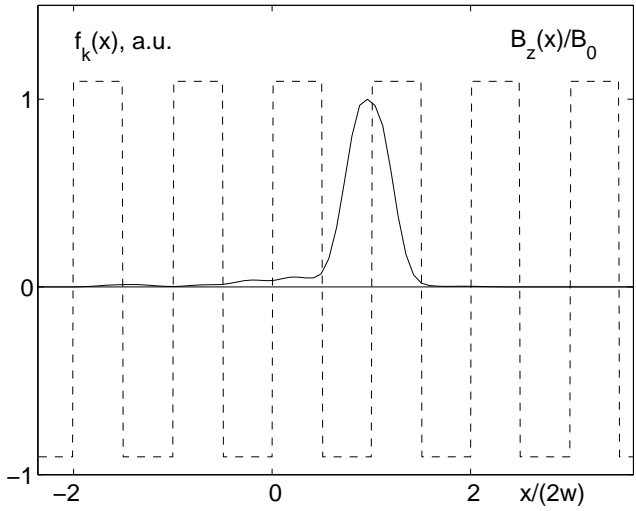}
 \end{center}
 \caption{\label{fig6}
The behavior of the ground state wavefunction (solid line) localized at a
domain wall in a periodic domain system. The magnetic field profile is
shown by the dash line. The parameters are $\pi B_0w^2/\Phi_0=5$ and
$H/B_0=0.095$.}
\end{figure}

For large values $w/L$ the phase transition line is very close to the one
found in the subsection \ref{2a}, except for the small temperature region
close to $T_{c0}$: $\Delta T\sim 4T_{c0} \xi_0^2/w^2$. Outside this narrow
temperature interval (and for $H<B_0$) the wavefunction is localized at
the domain walls (see Fig.\ref{fig6}).

\begin{figure}[htb]
 \begin{center}
 \epsfxsize=85mm
 \epsfbox{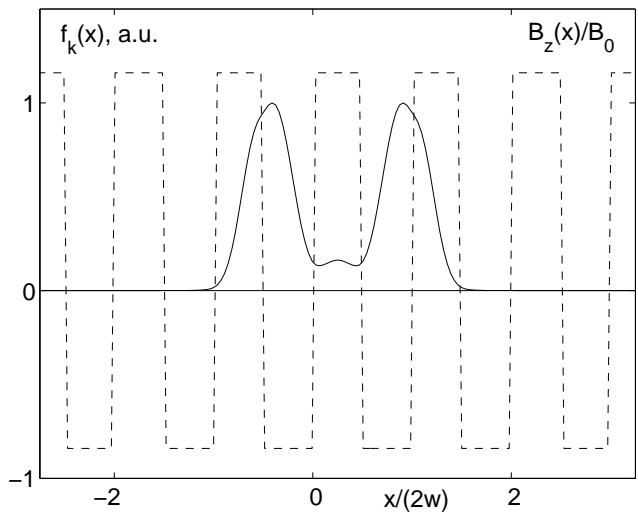}
 \end{center}
 \caption{\label{fig7}
The two peak structure of the ground state wavefunction (solid line) for a
periodic domain system. The magnetic field profile is shown by the dash
line. The parameters are $\pi B_0w^2/\Phi_0=5$ and $H/B_0=0.16$.}
\end{figure}

The coordinates of these localized nuclei shift at $ma$ as we change the
momentum at $4\pi H w m/\Phi_0$ ($m$ is an integer). Let us note, that for
rather weak magnetic fields $H<B_0$ we observed a very peculiar behavior
of the order parameter for a discrete set of field values given by the
condition $k^{min}_{2}-k^{min}_{1}=4\pi H w m/\Phi_0$: the ground state
wavefunction $f_k (x)$ has a two peak structure (see Fig.\ref{fig7}).
This fact is a natural consequence of the equivalence of the momenta $k$
and $k^\prime=k+ 4\pi H w m/\Phi_0$ and the resulting resonant interaction
of nuclei localized at domain walls separated by the distance $w(2m-1)$.

\begin{figure}[htb]
 \begin{center}
 \epsfxsize=85mm
 \epsfbox{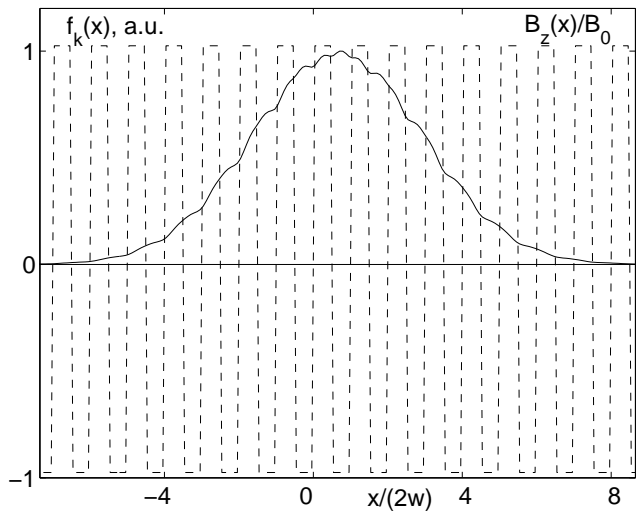}
 \end{center}
 \caption{\label{fig8}
The behavior of the ground state wavefunction in a periodic domain
structure for $\pi B_0w^2/\Phi_0=1$ and $H/B_0=0.025$ (solid line). The
magnetic field profile is shown by the dash line.}
\end{figure}

For not very large values $w/L< 2.0$ the critical temperature becomes a
monotonic function of the external magnetic field because of the strong
overlapping of wavefunctions corresponding to different domains.
Therefore, the wavefunction is no more localized in a single domain (see
Fig.\ref{fig8}).
However, even in this case we still observe a change in the slope of the
phase transition line (see dash line in Fig.\ref{fig5}).

The behavior of the upper critical field discussed above is not specific
for step-like field distributions. To demonstrate this fact we studied the
superconductivity nucleation for the field profile $B_z(x)=B_0\cos(2\pi
x/a)+H$. The phase diagram on the plane $H-T$ appears to be qualitatively
similar to the one shown in Fig.\ref{fig5}. The critical temperature is a
monotonic function of the external magnetic field for $a/L< 4.5$. For
large parameters $a/L\gg 1$ and $H<B_0$ the behavior of the critical
temperature can be analyzed analytically following the approach used in
subsection \ref{model}. The characteristic size of a superconducting
nucleus and the critical temperature of superconductivity nucleation are
given by the expressions:
\begin{eqnarray}
\label{Sin}
 \nonumber
 \ell = a\sqrt[3]{\frac{\Phi_0}{2\pi^2B_0a^2}}
 \left(1-\frac{H^2}{B_0^2}\right)^{-1/6},
 \\
 \left(1-\frac{T_c}{T_{c0}}\right)
 = \epsilon_0\frac{\xi^2_0}{a^2}
 \sqrt[3]{\frac{2\pi^2B_0a^2}{\Phi_0}\left(1-\frac{H^2}{B_0^2}\right)} \ .
\end{eqnarray}
The validity range of this approximate description is defined by the
conditions:
 $$
 \left(\frac{H/B_0}{1-H^2/B_0^2}\right)^{2/3}
 \ll\left(\frac{2\pi^2B_0a^2}{\Phi_0}\right)^{1/3}
 \quad{\rm and}\quad
 \left(\frac{2\pi^2B_0a^2}{\Phi_0}\right)^{1/3}\gg1 \ .
 $$

\section{Conclusions}
To summarize, we investigated the conditions of nucleation of localized
superconductivity at the domain boundaries in hybrid S/F systems. The
appearance of these localized superconducting nuclei should result in a
broadening of the superconducting transition probed, e.g., by the
resistivity measurements. We predict different regimes for the temperature
dependence of the upper critical field near $T_c$. The crossover between
these regimes could be easily seen on experiment. In fact, the beginning
of the resistivity decrease with the temperature decrease would correspond
to the domain wall superconductivity, while its complete disappearance
would signal the bulk superconductivity. External magnetic field would
shrink the region of the domain wall superconductivity. Let us discuss
some estimates of the physical parameters for the systems where the
nucleation of superconductivity at domain boundaries could be observable.
We can take, for example, the parameters of $Nb$ ($T_{c}\sim 9\ K$ and
$dH_{c2}/dT\sim 0.5\ kOe/K$) and typical values of magnetization for
ferromagnetic insulators $4\pi M\sim 1-10\ kOe$. The resulting increase in
the critical temperature above a domain wall is quite strong: $\delta
T_{c}\sim 1-3\ K $. The thickness of a superconducting film must be much
smaller than the distance between domains and ideal conditions correspond
to the thickness of the order of several coherence lengths. So we conclude
that the effects discussed above may be easily observed and could be quite
important. Note that the behavior observed in Ref. \cite{lange2} for S/F
bilayers with bubble domains in a ferromagnetic film is qualitatively
similar to our predictions. Generally the temperature behaviour of the
critical field in S/F structures can be very rich (see
Figs.\ref{fig2},\ref{fig3},\ref{fig5}) and it is strongly dependent on the
domain structure and method of determination of the critical field.
Careful measurements of the resistive and magnetic transition (including
the measurements of the transition broadening) on the samples with a
controllable domain structure would be very useful for the interpretation
of the phase diagram and could give an important information on the domain
wall superconductivity.

Note in conclusion, that the existence of localized superconducting
channels near the domain walls in S/F heterostructures can provide an
interesting possibility to realize a switching behavior provided we can
move the ferromagnetic domain wall. The superconducting channel in this
case should follow the motion of the domain wall, which provides a
possibility to control the conductance between certain static leads.

\section*{Acknowledgements}
We would like to thank Dr.~I.~A. Shereshevskii for fruitful discussions.
This work was supported, in part, by the Russian Foundation for Basic
Research, Grant No. 03-02-16774, Russian Academy of Sciences under the
Program 'Quantum Macrophysics', Russian State Fellowship for young doctors
of sciences (MD-141.2003.02), University of Nizhny Novgorod under the
program BRHE and 'Physics of Solid State Nanostructures', ESF 'Vortex'
Programme and Materials Theory Institute at the Argonne National
Laboratory.

\end{document}